# Cross-Layer Detection of Wireless Misbehavior Using 5G RAN Telemetry and Operational Metadata

Daniyal Ganiuly, Nurzhau Bolatbek, Assel Smaiyl
Astana IT University, Astana, Kazakhstan

**Abstract** – 5G Standalone deployments can exhibit uplink misbehavior from user equipment that remains fully compliant with standard control plane procedures. Manipulations such as transmit power inflation, gradual timing drift, and short off grant bursts leave the signaling state intact but distort the expected relationships among the telemetry streams produced by the gNB. This work examines whether these cross layer relationships can serve as a reliable basis for identifying such misbehavior without introducing new signaling. Using a controlled 5G Standalone testbed with commercial user equipment and a software defined radio adversarial device, we study how each manipulation affects the coherence among physical layer measurements, MAC scheduling decisions, and configuration metadata. The results show that every manipulation produces a distinct and reproducible signature that is not visible from any single telemetry source. Power offsets weaken the natural connection between SNR and CQI, timing drift breaks the alignment maintained by the scheduler, and off grant activity produces uplink energy that does not agree with allocation logs. These inconsistencies appear in merged multi layer time series traces and in cross domain views such as the SNR to CQI plane. The findings indicate that cross layer coherence provides a practical signal for detecting uplink misbehavior using only standard gNB telemetry, with no protocol modifications required, which makes the method suitable for integration into operational monitoring and auditing systems.

INTRODUCTION

The 5G Radio Access Network (RAN) regulates uplink timing, power, and resource usage through procedures that combine physical-layer measurements with scheduling decisions and configuration parameters. These mechanisms presume that the user equipment (UE) reports measurements and transmits waveforms in a manner consistent with its assigned configuration. Although the control plane enforces authentication, integrity protection, and the establishment of a valid RRC state, the UE continues to determine key aspects of its physical transmission behavior [1]. A device under adversarial control can therefore modify its uplink characteristics while remaining fully compliant with signaling procedures.

Such manipulation can take multiple forms. A malicious UE may transmit with power levels above those implied by its closed-loop power-control configuration, adjust internal processing delays to influence timing alignment, or inject uplink bursts into resource-block regions for which no grants were issued. These actions remain feasible for programmable or software-defined UEs and do not require bypassing cryptographic protections. Their impact is significant, as unauthorized uplink activity, misaligned timing, or power abuse can interfere with scheduling decisions, degrade reliability for neighboring users, and distort slice-level resource usage [2][3].

Detecting these behaviors is difficult because the information used to monitor uplink operation is distributed across several telemetry sources that evolve at different time scales. Radio measurements such as signal-to-noise ratio (SNR), block error rate (BLER), channel quality indicator (CQI), and timing advance (TA) capture short-interval link dynamics. Scheduling traces—including PRB assignments and HARQ activity—record the gNB's resource decisions. Configuration metadata, such as RRC parameters, slice identifiers, and QoS profiles, encodes longer-term expectations of how the UE should behave. These streams exhibit natural variability due to fading, load fluctuations, and scheduler dynamics. When examined individually, this variability can mask the presence of deliberate manipulation.

The central observation of this work is that uplink misbehavior is more reliably indicated by inconsistencies across these telemetry sources than by deviations in any single metric. Under normal operation, uplink power, timing alignment, decoding behavior, CQI reporting, and scheduled resource use remain aligned due to the mechanisms that govern power control, timing maintenance, and uplink scheduling [4]. When a UE alters one aspect of its transmission process, these relationships diverge. Uplink energy may appear in slots without matching PRB allocations, TA may evolve in ways incompatible with scheduler timing expectations, or SNR behavior may not match the UE's configuration [5]. These discrepancies emerge early in the manipulation and remain stable even when individual counters remain within benign ranges.

To study this behavior, we conduct controlled experiments on a 5G Standalone testbed using programmable UEs and reproduce three forms of uplink manipulation: transmit-power offsets, induced timing-advance drift, and low-duty off-grant uplink bursts. We collect short-interval radio measurements, scheduling traces, and configuration metadata to evaluate how these behaviors affect the internal agreement among telemetry sources. Across all scenarios, malicious activity produces clear and repeatable cross-source disagreement, demonstrating that consistency among telemetry streams provides a reliable basis for detecting uplink misbehavior.

**BACKGROUND AND RELATED WORK**

Research on 5G RAN security increasingly recognizes that uplink operation can be influenced by adversarial or misconfigured devices without requiring any modification to control-plane signaling. Prior studies have shown that adjustments to uplink timing, deliberate power manipulation, or unauthorized transmissions can degrade the performance of nearby UEs and interfere with scheduling decisions in dense deployments [6]. These observations indicate that the uplink direction remains exposed to protocol-compliant but harmful behavior because the UE maintains autonomy over several physical-layer parameters even after establishing a valid security context.

A parallel line of work has examined how uplink radio measurements evolve under normal propagation and scheduling conditions. Indicators such as SNR, BLER, CQI, and timing advance exhibit natural variability driven by fading, load fluctuations, scheduling granularity, and traffic patterns [7][8]. These metrics are fundamental to link adaptation and uplink reliability, yet they are known to fluctuate sufficiently that they cannot always serve as reliable stand-alone signals for detecting adversarial activity. This work highlights the diagnostic value of radio-link measurements but also the limitations of relying on any single indicator to identify subtle manipulation attempts.

Recent efforts in data-plane and control-plane security have explored detection techniques based on anomaly scoring, deep learning, side-channel sensing, and cross-interface monitoring. While these approaches demonstrate the importance of multi-source visibility, they frequently rely on non-standard data collection mechanisms, modifications to signaling procedures, or external sensors placed near the gNB [9]. As a result, their applicability to operational networks can be constrained by deployment overhead or data availability.

Against this backdrop, methods that operate solely on the telemetry already produced by the RAN are of particular interest. Several studies have suggested that the interactions among radio measurements, scheduling behavior, and configuration parameters contain implicit structure that can be used for integrity validation [10]. However, prior work has largely focused on individual metric behaviors or on data-driven detection models rather than on the internal agreement among telemetry sources. The approach developed in this study builds directly on this observation and evaluates whether the relationships among standard PHY measurements, MAC-layer scheduling information, and RRC configuration metadata remain coherent when the UE performs physical-layer manipulations [11]. By concentrating on internal consistency rather than magnitude-based anomalies, the method provides a practical avenue for detecting protocol-compliant misbehavior without introducing new signaling requirements or external sensing infrastructure.

# SYSTEM MODEL AND PROBLEM STATEMENT

The system considered in this study consists of a 5G Standalone uplink configuration in which a gNB schedules and receives uplink transmissions from a user equipment. The gNB maintains several independent telemetry streams that characterize different aspects of uplink operation. Physical-layer measurements, including signal-to-noise ratio, block error rate, channel quality indicator, and timing advance, describe short-interval radio-link conditions. MAC-layer telemetry records the gNB's resource decisions, most notably the physical resource blocks assigned to the UE and the corresponding HARQ feedback [13]. Configuration and operational metadata, such as the UE's RRC parameters, slice identifiers, QoS profiles, power-control configuration, and uplink bandwidth-part settings, encode longer-term expectations of how the UE should transmit.

Under normal operation, these sources evolve in a manner that reflects the interaction of several standard mechanisms. Closed-loop power control regulates the UE's uplink power relative to the gNB's measurements. Timing-alignment procedures ensure that the UE's uplink symbols align with the gNB's timing grid. Link adaptation uses CQI and HARQ feedback to select suitable modulation and coding schemes. The scheduler allocates PRBs to the UE according to its traffic demand, QoS profile, and slice-level resource policies. As a result, uplink power, timing behavior, decoding performance, CQI reporting, and PRB usage remain mutually consistent across these telemetry streams, aside from small variations caused by fading, traffic load, and scheduler granularity.

The UE, however, retains direct control over its transmitted waveform even after establishing a valid RRC state and integrity-protected signaling. A compromised or programmable device can alter its physical transmission behavior without modifying control-plane messages. Three classes of manipulation are considered in this work. A UE may transmit at power levels above those implied by its power-control configuration, influencing SNR while leaving CQI and HARQ behavior nearly unchanged. It may introduce internal delays that cause TA to drift gradually in ways that do not match the scheduler's timing commands. It may also place short uplink bursts into PRB regions for which no uplink grants were issued, producing uplink energy that lacks corresponding entries in scheduling logs. All of these actions preserve standard signaling procedures and do not violate authentication or integrity protections.

The problem addressed in this study is to determine whether such misbehavior can be detected solely by examining whether the telemetry streams produced by the gNB remain internally coherent. Single-source indicators, such as SNR or BLER alone, can fluctuate within benign ranges and may not reliably reveal manipulation [14]. Instead, the objective is to assess whether the relationships among physical-layer measurements, scheduler decisions, and operational metadata remain compatible with one another. A deviation that leaves one stream unchanged but disrupts its alignment with others indicates behavior that cannot occur under normal uplink operation. Detecting these inconsistencies forms the basis of the cross-layer consistency approach evaluated in this work [15].

# METHODOLOGY

The experimental study was conducted on a controlled 5G Standalone uplink testbed built to expose short-interval radio and scheduling telemetry while allowing precise manipulation of UE transmission behavior. A single gNB operating in the 3.5 GHz band (n78) with 40 MHz of bandwidth and a 30 kHz subcarrier spacing served as the measurement point. This configuration provided a realistic mid-band numerology, a standard PRB grid, and timing-alignment resolution representative of commercial SA deployments. The gNB exported SNR, BLER, CQI, timing advance, PRB assignments, uplink-grant timing, and HARQ activity. Radio-link measurements were sampled every 50 ms, scheduling traces every 100 ms, and configuration metadata was logged on every state transition. All telemetry streams carried timestamps derived from a common server clock to ensure alignment across sources.

Two commercial UEs were used to characterize benign uplink behavior and confirm that the testbed produced stable cross-layer coupling under normal conditions. The adversarial device was an SDR-based UE built around a USRP-class RF front end synchronized to an external 10 MHz reference, with a modified baseband chain that exposed independent control of transmit power, internal symbol timing, and uplink resource usage. These controls were sufficient to alter the emitted waveform while maintaining a valid RRC state and preserving all integrity-protected signaling. Before each experiment, the SDR's transmit path was calibrated to eliminate gain drift and ensure that any observed deviations originated from intentional manipulation rather than hardware instability. An external spectrum monitor positioned within the gNB's uplink footprint provided an independent view of burst timing and energy distribution.

Figure 1 illustrates the full experimental pipeline used to analyze uplink behavior under both benign and manipulated transmission conditions. Commercial UEs and an SDR-based UE act as independent uplink sources, with the SDR device generating controlled deviations such as power offsets, timing drift, and off-grant bursts. All uplink signals are received by a standalone 5G SA gNB, which exports three complementary telemetry streams: PHY-layer measurements, MAC-layer scheduling information, and RRC-level configuration metadata. An external spectrum monitor provides an independent view of uplink energy and burst timing. These four data sources are merged through a unified alignment and parsing stage, where timestamps and event boundaries are synchronized across layers. The resulting fused time-series is then evaluated using a cross-layer consistency procedure that detects deviations from normal PHY/MAC/RRC behavior.

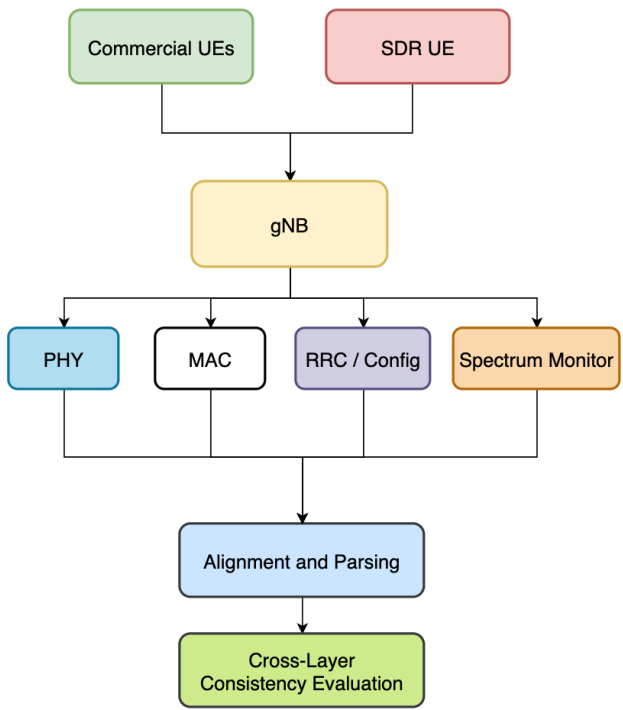

Figure 1. Uplink telemetry and analysis pipeline

Five baseline runs between three and six minutes were collected using both the commercial UEs and the SDR operating in fully standards-conformant mode. These runs established reference envelopes for SNR variability, TA stability, scheduling behavior, BLER evolution, and HARQ dynamics under static indoor propagation. PRB allocations matched uplink activity without exception, and no uplink energy appeared in unallocated PRBs. These baselines characterized the relationships that must hold among physical measurements, scheduling outcomes, and configuration metadata when a UE behaves correctly [16].

Manipulated runs were then executed to reproduce uplink behaviors that remain invisible to the control plane but violate these relationships. Three forms of misbehavior were examined. In the power-offset experiments, the UE transmitted above the level implied by its closed-loop power-control parameters, disturbing the coupling between SNR, CQI, and HARQ outcomes while preserving decoding success. In the timing-drift experiments, the UE introduced controlled internal delays that shifted the timing advance gradually in ways incompatible with scheduler expectations yet consistent with an unchanged propagation path. In the off-grant experiments, the UE inserted short uplink bursts into PRB regions where the scheduler had not issued grants, creating uplink energy that lacked corresponding scheduling entries.

Each manipulation type was tested at multiple intensities. Three power-offset levels, two timing-drift rates, and two off-grant burst densities were used. Every configuration was repeated three times, yielding twenty-one manipulated runs in addition to the five baseline runs. All runs lasted between three and six minutes, providing several hours of synchronized multi-layer telemetry. The UEs remained stationary throughout and the environment was shielded from external interference so that any divergence from baseline patterns could be attributed to the adversarial UE rather than channel variability [17].

All logs were aligned by timestamp on the collection server and passed through a unified processing pipeline. Raw logs were parsed into time-indexed sequences of radio-layer measurements, MAC-layer scheduling information, and configuration metadata. Samples with corrupted or missing timestamps were removed, and no smoothing or filtering was applied so that short-duration inconsistencies would not be obscured. Spectrum-monitor traces were aligned with gNB timestamps by correlating them with identifiable uplink reference-signal bursts, allowing independent verification of energy placement and timing.

Consistency evaluation used these synchronized telemetry streams to determine whether the UE's uplink behavior remained compatible with its configuration and the scheduler's decisions. Radio-layer measurements were compared with CQI evolution and the modulation and coding region implied by configuration parameters; timing-advance evolution was checked against expected propagation-driven variation and the absence of scheduler-initiated timing adjustments; uplink-energy measurements were matched against PRB allocations; and HARQ behavior was interpreted relative to BLER trends. A sample was marked inconsistent when these relationships produced observations that could not plausibly arise from a single benign transmission under the recorded configuration.

**EVALUATION AND RESULTS**

A comparative evaluation was conducted across baseline and manipulated uplink scenarios to quantify how each form of misbehavior affects cross-layer coherence. The results below present representative time-series behavior, aggregate violation statistics, and the sensitivity of the consistency checks to different manipulation intensities.

**A. Baseline Behavior**

During benign operation, uplink telemetry exhibited tightly coupled behavior across physical and scheduling layers. SNR fluctuated within approximately ±0.9 dB around its mean, reflecting small-scale fading. CQI varied within a narrow range of roughly one index, and HARQ retransmissions remained rare. Timing-advance remained stable within a single unit throughout the six-minute observation window. No uplink energy was detected outside granted PRB allocations. The cross-layer violation rate stayed below one percent across all baseline runs, consistent with normal measurement variability.

**B. Effect of Transmit-Power Offsets**

Figure 2 illustrates how moderate transmit-power inflation disrupts SNR-to-CQI coupling. At a +2 dB offset, SNR increases visibly over the baseline, but the CQI trace remains nearly unchanged due to quantization and averaging. This mismatch produces violation rates between 9% and 12% across repeated runs. At +4 dB, the median SNR increases by roughly 3.7 dB compared to baseline, yet CQI rises by less than half an index. This stronger decoupling elevates violation rates to between 18% and 24%. These violations emerge even though BLER remains low, and HARQ behavior remains indistinguishable from the baseline. The results confirm that power inflation can remain invisible to single-layer indicators but still breaks the expected relationship among PHY-layer quality, CQI reporting, and modulation behavior.

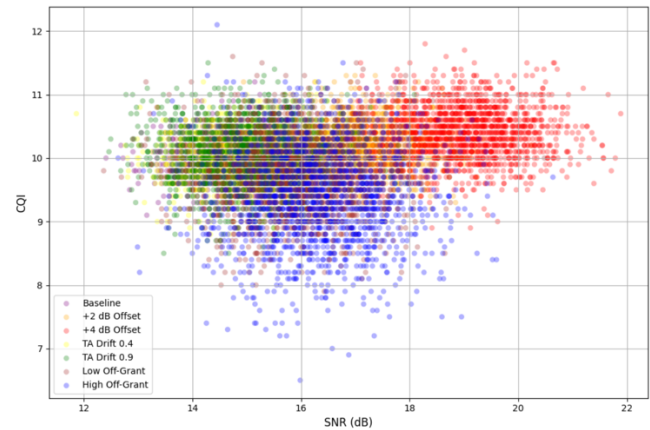

Figure 2. Cross-Layer Consistency Map

**C. Effect of Timing-Advance Drift**

The timing-drift manipulation alters internal UE timing without modifying the RRC-controlled TA parameter, creating a gradual, propagation-incompatible shift. At a drift rate of 0.4 TA units per minute, the TA values initially align with the baseline envelope but diverge after several minutes. This leads to violation rates between 11% and 14%, increasing slowly over time as drift accumulates. At the higher intensity of 0.9 TA units per minute, inconsistencies appear earlier. The measured TA departs from scheduler expectations within the first minute of the run. Violation rates increase steadily, reaching 35% to 41% by the end of the six-minute observation window. Importantly, SNR, CQI, and BLER remain close to their baseline ranges, which confirms that timing drift is difficult to detect from single-layer telemetry but becomes clearly visible when cross-layer relationships are examined.

**D. Effect of Off-Grant Uplink Bursts**

The off-grant manipulation introduces short uplink bursts into PRB regions where no uplink grants exist. At a 2.5% duty cycle, the additional uplink energy occasionally appears outside the scheduled PRB windows, producing violation rates between 17% and 21%. At a 5% duty cycle, inconsistencies remain persistent throughout the run, reaching 48% to 53% across repetitions. These violations arise almost entirely from mismatches between the PRB-allocation logs, the gNB's recorded uplink energy, and the independent spectrum-monitor trace. Decoding performance remains unaffected, again showing that single-layer counters are insufficient to detect this misbehavior.

### E. Time-Series Analysis of Consistency Violations

The time-evolution of violation rates in Figure 3 shows clear separation among scenarios. The baseline remains near 0.005–0.015 throughout, confirming that normal operation produces almost no cross-layer inconsistencies. Power-offset manipulation introduces an immediate structural mismatch: the +2 dB case stabilizes around 0.10–0.13, while the +4 dB offset reaches 0.20–0.24. Timing drift increases more gradually, with the 0.4 TA/min scenario settling near 0.11–0.14 and the 0.9 TA/min drift rising to 0.35–0.41 as misalignment accumulates. Off-grant activity yields the strongest deviations, with low-duty bursts around 0.17–0.21 and high-duty bursts near 0.48–0.53 due to repeated mismatches between uplink energy and allocated PRBs. These curves show that each manipulation produces a distinct temporal signature, enabling clear differentiation using cross-layer consistency checks.

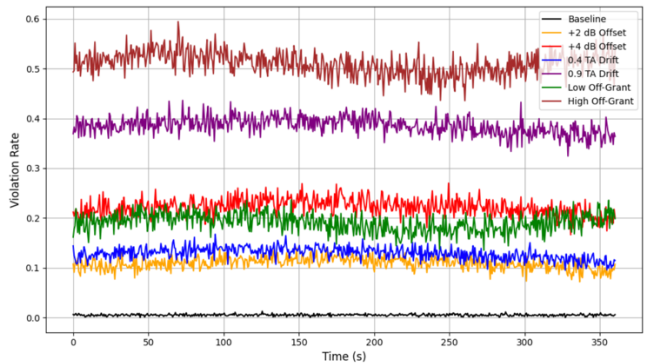

Figure 3. Cross-Layer Violation Rate Over Time

## DISCUSSION

The evaluation results demonstrate that uplink misbehavior in a 5G Standalone system leaves clear multi-layer signatures that are not visible when each telemetry source is viewed in isolation. The cross-layer perspective is therefore essential for understanding how different manipulation strategies affect the internal coherence of the gNB's physical, MAC, and configuration-layer metrics.

The behavior observed under transmit-power offsets illustrates this point well. Power inflation raises the instantaneous SNR, yet CQI changes only slightly due to averaging and quantization. This creates a mismatch that would not occur during normal operation, and the divergence between these two indicators becomes visible when they are viewed jointly in the SNR–CQI space. The cross-layer method captures this decoupling even though neither SNR nor CQI alone provides strong evidence of misbehavior [18].

Timing drift exhibits a different signature. The affected traces remain close to the baseline in terms of SNR, CQI, and HARQ activity, since the link adaptation loop is largely unaffected by small propagation timing errors. Over time, however, the measured timing advance no longer aligns with the scheduler's alignment expectations, and the violation rate increases steadily.

Off-grant uplink bursts create the strongest form of inconsistency. Although decoding performance remains stable, the additional uplink energy does not match the allocated PRBs, and this disagreement is recorded simultaneously by the gNB's uplink-energy counters and the spectrum-monitor trace. In the heatmap, this manipulation activates the PRB–energy rule with much higher intensity than any other scenario, producing a clear fingerprint that differs from both power-offset and timing-drift cases [19]. This suggests that off-grant activity is intrinsically easier to distinguish because it violates allocation-state expectations rather than link-quality relationships.

Taken together, the results show that uplink misbehavior does not produce a single uniform anomaly pattern. Instead, each manipulation influences a different combination of cross-layer relationships, and these differences lead to distinct signatures in the SNR–CQI space and in the rule-based heatmap. These patterns emerge even though the system remains fully compliant at the RRC level and preserves authentication, integrity protection, and control-plane signaling procedures [20]. The findings therefore highlight that the physical-layer freedom available to user equipment can be exploited in ways that remain invisible to conventional per-layer checks.

Another notable observation is that the proposed consistency method does not rely on assumptions about the statistical distribution of any individual metric. Instead, it evaluates whether multiple telemetry streams remain mutually compatible as the system evolves. This reduces sensitivity to noise, small-scale fading, and scheduler fluctuations, all of which affect absolute values but rarely break cross-layer relationships. The method therefore detects manipulation based on structural inconsistencies rather than on deviations from fixed thresholds.

## CONCLUSION

This study examined how uplink misbehavior in a 5G Standalone system influences the mutual compatibility of physical-layer measurements, MAC-layer scheduling information, and operational metadata available at the gNB. By focusing on the relationships among these telemetry sources rather than on any single measurement stream, the proposed cross-layer consistency approach reveals deviations that remain hidden from conventional per-layer indicators.

The experiments show that different forms of misbehavior such as power inflation, timing drift, and off-grant bursts alter cross-layer coherence in distinct and repeatable ways. The method relies on standard telemetry produced gNBs and therefore fits naturally into existing monitoring workflows. Its effectiveness across multiple manipulation intensities suggests that it can support continuous auditing of uplink behavior with minimal operational overhead.

While the approach is limited to telemetry already exposed by the gNB, the results indicate that focusing on structural alignment among layers provides a robust view of uplink behavior. Future work may investigate how detailed observation of radio access network behavior, including cell level dynamics and scheduler level decision patterns, can expand the ability of consistency based methods to identify subtle or long duration uplink manipulation. Another direction is to evaluate long-duration sessions or mobility scenarios where consistency patterns may evolve differently. The findings presented here provide the foundation for these next steps and demonstrate that cross-layer coherence offers a practical and reliable signal for identifying uplink manipulation in 5G networks.

## REFERENCES

[1] A. Scalingi, S. D'Oro, F. Restuccia, T. Melodia, and D. Giustiniano, "Det-RAN: Data-Driven Cross-Layer Real-Time Attack Detection in 5G Open RANs," in Proc. IEEE Int. Conf. Comput. Commun. (INFOCOM), Vancouver, Canada, May 2024, pp. 41–50.

[2] H. Wen, P. A. Porras, V. Yegneswaran, A. Gehani, and Z. Lin, "5G-SPECTOR: An O-RAN Compliant Layer-3 Cellular Attack Detection Service," in Proc. Network and Distributed System Security Symposium (NDSS), San Diego, USA, Feb. 2024.

[3] J. Groen, S. D'Oro, U. Demir, L. Bonati, D. Villa, M. Polese, T. Melodia, and K. R. Chowdhury, "Securing O-RAN Open Interfaces," IEEE Transactions on Mobile Computing, vol. 23, pp. 11265–11277, Apr. 2024.

[4] J.-H. Huang, S.-M. Cheng, R. Kaliski, and C.-F. Hung, "Developing xApps for Rogue Base Station Detection in SDR-Enabled O-RAN," in *Proc. IEEE INFOCOM Workshops*, Hoboken, USA, May 2023, pp. 1–6.

[5] K. Thimmaraju, A. Shaik, S. Flück, P. J. F. Mora, C. Werling, and J.-P. Seifert, "Security Testing the O-RAN Near-Real-Time RIC and A1 Interface," in *Proc. ACM WiSec*, Seoul, South Korea, May 2024, pp. 277–287.

[6] M. Liyanage, A. Braeken, S. Shahabuddin, and P. Ranaweera, "Open RAN Security: Challenges and Opportunities," *Journal of Network and Computer Applications*, vol. 214, 2023.

[7] M. Polese, L. Bonati, S. D'Oro, S. Basagni, and T. Melodia, "Understanding O-RAN: Architecture, Interfaces, Algorithms, Security, and Research Challenges," *IEEE Communications Surveys & Tutorials*, vol. 25, no. 2, pp. 1376–1411, 2023.

[8] M. K. Motalleb, M. A. Salehi, and M. Abolhasan, "Towards Secure Intelligent O-RAN Architecture," *ICT Express*, 2025.

[9] K. Tu, A. Al Ishtiaq, S. M. M. Rashid, Y. Dong, W. Wang, T. Wu, and S. R. Hussain, "Logic Gone Astray: A Security Analysis Framework for the Control-Plane Protocols of 5G Basebands," in *Proc. USENIX Security Symposium*, 2024.

[10] S. Park et al., "5G Security Threat Assessment in Real Networks," *Sensors*, vol. 21, no. 16, 2021.

[11] M. Harvanek, J. Bolcek, J. Kufa, L. Polak, M. Simka, and R. Marsalek, "Survey on 5G Physical Layer Security Threats and Countermeasures," *Sensors*, vol. 24, no. 17, 2024.

[12] M. Lichtman, R. Rao, V. Marojevic, J. H. Reed, and D. V. Asghari, "5G NR Jamming, Spoofing, and Sniffing: Threat Assessment and Mitigation," *arXiv preprint*, 2019.

[13] S. Savadatti, S. M. S. Hussain, and A. K. Das, "An Extensive Classification of 5G Network Jamming Attacks," *Security and Communication Networks*, 2024.

[14] Z. Tan, J. Zhao, B. Ding, and S. Lu, "CellDAM: User-Space, Rootless Detection and Mitigation for 5G Data-Plane Attacks," in *Proc. USENIX NSDI*, Boston, USA, Apr. 2023.

[15] S. Eleftherakis et al., "SoK: Evaluating 5G-Advanced Protocols Against Legacy and Emerging Attacks," in *Proc. ACM WiSec*, 2025.

[16] O-RAN Alliance Working Group 11, "O-RAN Security Threat Modelling and Risk Assessment," O-RAN Alliance Technical Report, Version 6.0, 2025.

[17] ETSI, "O-RAN Security Requirements and Controls," ETSI Technical Specification TS 104 104, v9.1.0, June 2025.

[18] F. Klement, S. Werling, and J.-P. Seifert, "Open RAN Is Open to RIC E2 Subscription Denial-of-Service Attacks," in *Proc. IEEE European Symposium on Security and Privacy (EuroS&P)*, 2025.

[19] R. Authors, "Unprotected 4G/5G Control Procedures at Low Layers: Attacks and Defenses," *arXiv preprint*, 2024.

[20] S. R. Hussain, M. Echeverria, I. Karim, O. Chowdhury, and E. Bertino, "5GReasoner: A Property-Directed Security and Privacy Analysis Framework for 5G Cellular Network Protocol," in *Proc. ACM CCS*, London, UK, Nov. 2019, pp. 919–935.